# Evidence for High Eccentricity and Apsidal Motion in the Detached Eclipsing Binary GSC 04052-01378

**Riccardo Furgoni**
*Keyhole Observatory MPC K48, Via Fossamana 86, San Giorgio di Mantova (MN), Italy; riccardo.furgoni@gmail.com*

**Gary Billings**
*Monument Hill Observatory, P.O. Box 263, Rockyford, AB T0J 2R0, Canada; obs681@gmail.com*

**Abstract**   We observed the recently discovered eccentric eclipsing binary GSC 04052-01378 in order to improve the light curve parameters and to find further evidence of its probable high eccentricity and of apsidal motion. Furthermore, we propose a basic stellar model that fits very well the observations of our dataset and where the eccentricity is found to be e = 0.538(6), which corresponds to an high value among this group of binaries.

### 1. Eccentric detached eclipsing binaries: some considerations about this type of system and serendipitous photometric discovery

Detached eclipsing binary systems (β Persei or "EA" type) are one of the largest groups in any catalog of variable stars. Both in the *General Catalogue of Variable Stars* (GCVS; Samus *et al.* 2017), and in the more recently created International Variable Star Index (VSX; Watson *et al.* 2014), there are thousands of members and at least a few hundred more that are strongly suspected to be of this type. This is not surprising: stellar multiplicity (systems with two or more components) is common, with the frequency declining with the number of components present (Tokovinin 2014).

Whether single star systems are more common than multiple systems is still unresolved. Mathieu (1994, p. 517) in his careful analysis of the stellar multiplicity in the pre-main sequence stars, said that the formation of binary stars is the primary branch of the star-forming processes. More recently Lada (2006) observed on the contrary that "most stellar systems formed in the Galaxy are likely single and not binary, as has been often asserted. Indeed, in the current epoch two-thirds of all main-sequence stellar systems in the Galactic disk are composed of single stars."

That this question is unresolved is probably associated with the lack of a definitive theory for the formation of multiple systems. According to Tohline (2002), one of the most promising theories explaining formation of multiple systems is the fragmentation of the pre-stellar core in the early stages of its collapse, controlled by factors such as pressure, rotation, turbulence, and magnetic fields (Commerçon *et al.* 2010). Alternately, Bonnel (1994) presents the option of early fragmentation of a circumstellar disk, especially when the mass of the disk itself is higher than that of the protostar in the early stages of its formation.

In a recent review of the literature about this subject, Duchêne and Kraus (2013) concluded that the degree of multiplicity is directly related to the mass of the primary component, according to the results shown in Table 1.

Table 1. Multiplicity properties for Population I main sequence stars and field brown dwarfs (Duchêne and Kraus 2013).

| *Category* | *Mass Range* $(M_\odot)$ | *Multiple System / Companion Frequency* |
|---|---|---|
| VLM/BD | $\lesssim 0.1$ | MF = 22 % <br> CF = 22 % |
| M | 0.1–0.5 | MF = 26 ± 3% <br> CF = 33 ± 5% |
| FGK | 0.7–1.3 | MF = 44 ± 2% <br> CF = 62 ± 3% |
| A | 1.5–5 | MF ≥ 50% <br> CF = 100 ± 10% |
| Early B | 8–16 | MF ≥ 60% <br> CF = 100 ± 20% |
| O | $\gtrsim 16$ | MF ≥ 80% <br> CF = 130 ± 20% |

Note: MF is the frequency of multiple systems and CF the companion frequency, whereas it must be noted that this last can exceed 100%.

All these data justify the large number of detached eclipsing systems in variable star catalogs, and lead us to examine the nature of the orbits of those systems, in particular systems showing a marked orbital eccentricity.

Only a small fraction of known detached eclipsing binary systems have a detectable eccentricity, while for most the eccentricity is near zero (essentially circular orbits). Amongst short period systems (more easily discovered and studied) it is most likely that the orbit has already been circularized by tidal interactions.

Since tidal interactions will circularize orbits that were initially more eccentric, we should expect to see greater eccentricity, and more frequent occurrence of eccentric systems, amongst younger stellar systems, and systems of "earlier" spectral type (see on this subject Mazeh 2008).

To determine the orbital radii (and the related quantity, orbital period) corresponding to circularized orbits is not straightforward. But empirically, orbits are usually circularized in systems with periods shorter than ~7.1 days for pre-main sequence stars with an age between ~1 and ~10 Myr, up to

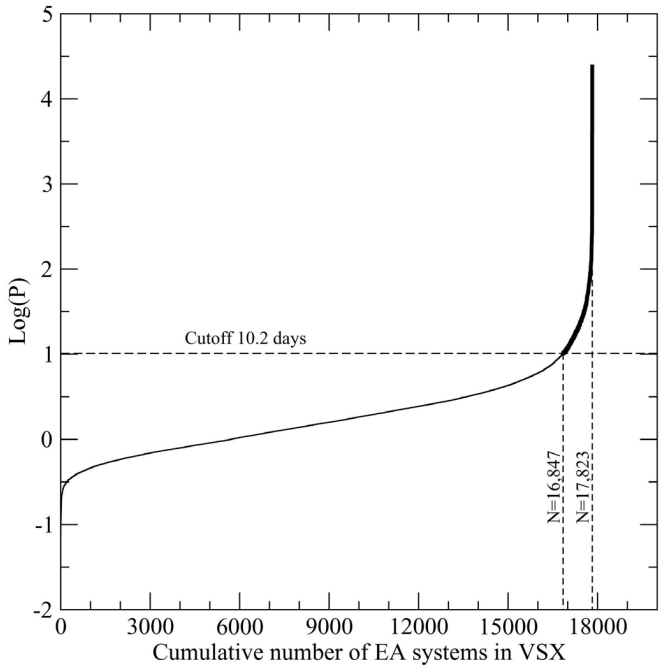

Figure 1. The graph shows the cumulative number of EA systems in VSX with specified period (excluding systems with unknown or uncertain period) in relation to the logarithm of the orbital period. The horizontal dashed line represents 10.2 days, the value of the circularization period used for this analysis.

~15.6 days for the stars of the Galactic Halo with estimated age of ~10 Gyr (Meibom and Mathieu 2005). For main sequence stars (in M35), the same authors showed that orbits with periods shorter than ~10.2 days are circularized. We apply this latter period as a cutoff to the entire population of detached binaries found in VSX to get an estimate the population of the number of eccentric (non-circularized) systems. The result is shown in Figure 1.

Of the 17,823 EA systems listed in VSX with a defined orbital period, only 976 (5.47%) exceed our circularization cutoff period of 10.2 days. No correction has been attempted for selection effects (longer period systems are less likely to be discovered). Thus, we expect eccentric systems to be rare.

## 2. GSC 04052-01378

This binary system was discovered by R. Furgoni and S. Otero and then added to VSX on April 4, 2014. Subsequently, complete information about the system has appeared in Furgoni (2014). The variable was classified as a β Persei (EA) after evaluation of the phase plot. The basic data of this system as well as the stars used for comparison in the paper published by Furgoni (2014) are presented in Table 2.

The system was immediately identified as eccentric, and with probable apsidal motion. The eccentricity was recognized by the large difference in duration between primary and secondary eclipses while the apsidal motion was suggested by a better matching of the secondary eclipse observations using a slightly shorter period than for primary eclipses. Both conclusions can be assessed by examining the data originally published in Furgoni (2014).

The interesting difference in the primary and secondary eclipse durations, where the secondary is more than three times longer than the primary, led to the planning of a new observational campaign aimed at a better characterization of the light curve for subsequent modeling.

## 3. Instrumentation used and observation details

The observations were made in two observatories at a great distance in longitude corresponding to a time difference of 8 hours. This was advantageous to observing more eclipses, in light of the long period and the period being a non-integer number of days.

At the observatory managed by R. Furgoni, two different telescopes were used. The main instrument was a TS Optics APO906 Carbon apochromatic refractor with 90 mm aperture and f/6.6 focal ratio, and in the last observational sessions, a Celestron C8 Starbright Schmidt-Cassegrain with aperture of 203 mm and Baader Planetarium Alan Gee II focal reducer, yielding f/6. With both telescopes photometry was done with a CCD SBIG ST8300m equipped with a Kodak KAF8300m monochromatic sensor. The Johnson V passband photometry was performed with an Astrodon Photometrics Johnson-V 50 mm round filter.

At the observatory managed by G. Billings, observations were made using a Celestron C-14 (14" f/11 Schmidt-Cassegrain) telescope and an SBIG STL-6303E CCD camera and a Bessell-prescription Johnson V filter. Image processing was conventional dark subtraction and flat fielding, performed using Starlink software (Currie *et al.* 2014). Photometry was performed using the Starlink operation "photom" with measurement apertures that varied night-to-night depending on the seeing-limited PSF.

The log of the authors' observations used in the graphs and analysis is presented in Table 3.

Table 2. Position, identification, and light elements of GSC 04052-01378 as presented in Furgoni (2014).

| | |
|---|---|
| Position (UCAC4)[1] | R.A. (J2000) = $02^h 53^m 08.34^s$, Dec. (J2000) = +62° 06' 10.5" |
| Cross Identification | UCAC4 761-021922; NSVS 1888562; 1SWASP J025308.36+620610.7 |
| Variability Type | EA |
| Magnitude Range | Max. = 11.76 V, Min. = 12.08: V |
| Spectral type | B2 |
| Period | 18.3024(1) d |
| Epoch | 2451403.83(1) HJD |
| Ensemble Comparison Stars | UCAC4 761-021906 (APASS 12.498 V); UCAC 4 761-021905 (APASS 12.566 V) |
| Check Star | UCAC4 761-022036 |

*1. Zacharias et al. 2012.*

Table 3. Log of observations used in the analysis and that were fit by modeling.

| Observer | RJD[1] | Start Time (UT) | End Time (UT) | Telescope | Notes |
|---|---|---|---|---|---|
| RF | 6630 | 03/12/2013 17:28 | 03/12/2013 23:53 | Celestron 8 | Non-eclipse |
| RF | 6631 | 04/12/2013 17:17 | 04/12/2013 22:43 | Celestron 8 | Non-eclipse |
| RF | 6632 | 05/12/2013 17:03 | 05/12/2013 21:39 | Celestron 8 | Non-eclipse |
| RF | 6706 | 17/02/2014 17:43 | 17/02/2014 21:45 | Celestron 8 | Non-eclipse |
| RF | 6958 | 27/10/2014 19:42 | 27/10/2014 23:28 | TS 906 APO | (s) ingress |
| RF | 6959 | 28/10/2014 17:15 | 28/10/2014 22:11 | TS 906 APO | (s) egress |
| RF | 7022 | 30/12/2014 22:09 | 31/12/2014 00:50 | TS 906 APO | (p) ingress |
| RF | 7114 | 01/04/2015 19:13 | 01/04/2015 21:25 | TS 906 APO | (p) egress |
| GB | 7699 | 07/11/2016 03:36 | 07/11/2016 13:59 | Celestron 14 | (p) ingress and egress |
| GB | 7702 | 10/11/2016 05:38 | 10/11/2016 06:17 | Celestron 14 | Non-eclipse |
| GB | 7725 | 03/12/2016 07:05 | 03/12/2016 07:31 | Celestron 14 | Non-eclipse |
| GB | 7726 | 04/12/2016 05:57 | 04/12/2016 09:02 | Celestron 14 | (s) ingress |
| GB | 7745 | 23/12/2016 01:45 | 23/12/2016 12:48 | Celestron 14 | (s) egress |
| GB | 7781 | 28/01/2017 06:25 | 28/01/2017 13:56 | Celestron 14 | (s) ingress |
| GB | 7782 | 29/01/2017 04:29 | 29/01/2017 08:00 | Celestron 14 | (s) egress |

*1. Reduced Julian date, viz., JD – 2450000.*

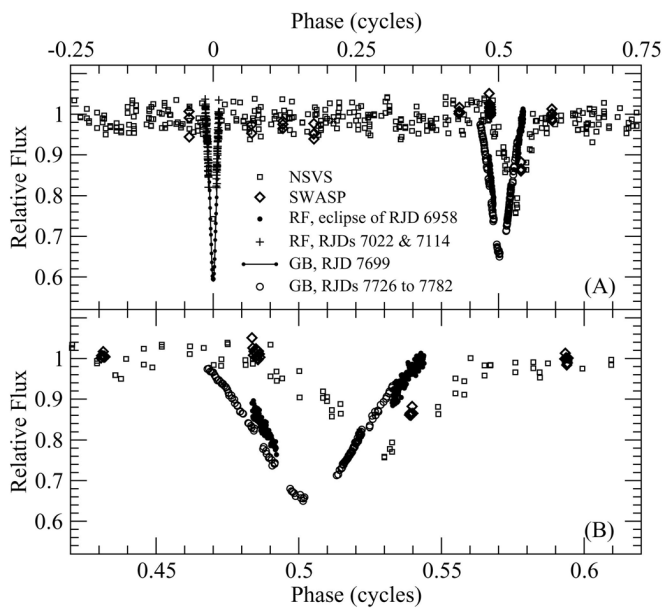

Figure 2. The upper panel (A) shows the full cycle phased light curve using elements from Furgoni (2014). The NSVS data, which span the whole cycle, are relatively noisy but show no out-of-eclipse variation. The lower panel (B) is an enlargement of the secondary eclipse. The different datasets have minima at different phases. Each data source is clustered around a different epoch, so each is showing the secondary eclipse at a different point in the rotation of the apse. The difference between the NSVS data and the observations by GB of RJD 7726 - 7782 also suggests a different eclipse depth, but this is not certain as the two datasets are not transformed to a common photometric system.

For this star we made use of observations obtained by the Northern Sky Variability Survey (NSVS) (Wozniak *et al.* 2004) and by the Wide Angle Search for Planets (SuperWASP) (Butters *et al.* 2010), in addition to those made by the authors at their private observatories. The authors' data are untransformed V-filtered differential aperture photometry. The authors' data are attributed using the authors' intials (RF or GB), and the time series are referred to using reduced Julian data (RJD), that is, the last four digits of the Julian date, 7699, for example, for the time series from Julian date 2457699.

RF's differential photometry was performed using ensemble photometry as described in Table 2 and Furgoni (2014). GB used single comparison and check stars (GSC 4052-1048 and 4052-0634) selected for having similar color to the target, as determined from the AAVSO Photometric All-Sky Survey (APASS DR9; Henden *et al.* 2015).

These datasets were not inter-calibrated or transformed to a standard system, except for zeropoint shifts determined by graphical comparison. Only one, constant, shift was used for each data source. Thus, inter-night zero-point offsets from each observer were not removed. Visual inspection of the light curves (at larger scale) suggests such offsets are occasionally present at the level of as much as a few percent. This level of uncertainty inevitably limits the precision and accuracy of the modeling that follows.

Finally, each dataset was converted from magnitude to flux, so the light curves could be displayed along with the modeling results in the program BINARYMAKER3 (Bradstreet and Steelman 2004).

The data used here have been deposited in the AAVSO International Database (Kafka 2017; observer codes FRIC and BGW, and star AUID 000-BLH-415).

### 4. Analysis and modeling

Figure 2(A) is the phased light curve for this star. It shows the NSVS and SWASP data used in Furgoni's earlier paper, as well as new time-series concentrated on the minima, taken by the present authors. All the data in Figure 2 are "phased" using elements that give a best fit to the times of all primary eclipses (period 18.3024 d). These data confirm the dramatically different eclipse widths noted by Furgoni (2014), and show that the narrow eclipse is the primary (deepest) eclipse.

Figure 2(B). An enlargement of the data around the secondary eclipse. The different datasets are clustered in time around different dates. The NSVS data ranges from RJD 1370 to 1609 (midpoint RJD 1490), and SWASP data from 3196 to 3226 (midpoint RJD 3212). Furgoni (2014) determined a time of (secondary) minimum from these data, with the NSVS data dominating the result. RF's data from around RJD 6958 (about

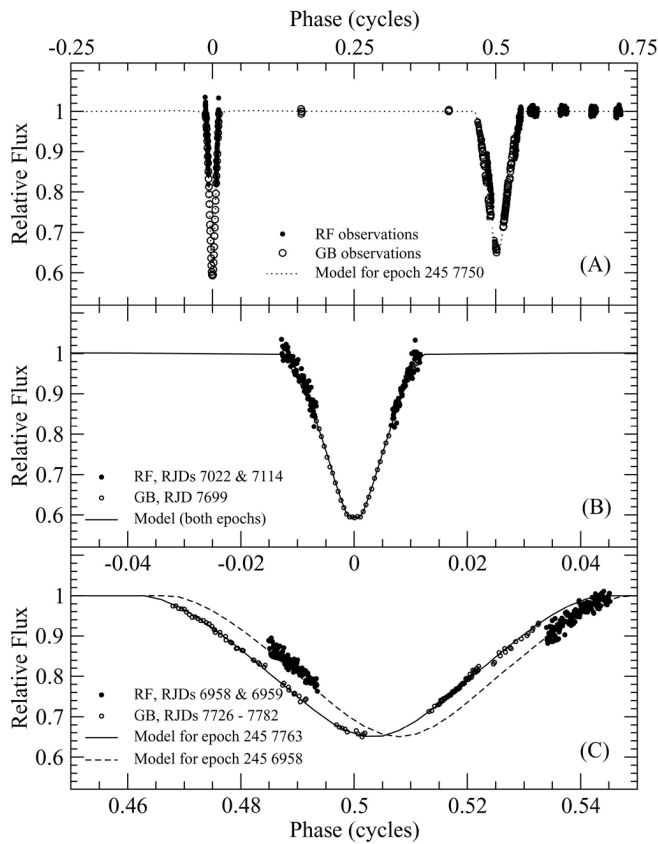

Figure 3: Data used in fitting a model to the observations, and the resulting modeled light curve. The upper panel (A) shows the full cycle. Several sets of observations from out-of-eclipse are shown; these were used to establish the relative zeropoints between the observers, and to set the level for flux=1.0. The middle panel (B) is an enlargement of the primary eclipse. The elements used to phase all three panels were chosen to perfectly phase the primary eclipse in the GB and RF datasets (from two epochs, about two years apart). The lower panel (C) is an enlargement of the secondary eclipse. The horizontal scale, but not its range, is the same as for the middle panel, to demonstrate the difference in eclipse widths. The GB and RF datasets are each from a short range of dates about 2 years apart, and show a distinctly different phase for the eclipse. A second model (dashed line) was created so both datasets could be fit; the only difference between the models is longitude of perihelion, viz., rotation of the apse.

15 years later than the NSVS observations) come at an earlier phase, and GB's data from around 7763 (about 2 years later still) comes even earlier. Thus, the secondary eclipse is consistently shifting to be closer in time to the preceding primary eclipse, indicating apsidal rotation.

The data during the secondary eclipse also suggest that the eclipse depth is changing (see Guilbault *et al.* 2001 for an example of another system showing changing eclipse depths). However, this is not entirely proven because the datasets are not inter-calibrated.

In the following, we describe the steps used to "model" the light curve. This type of model is not merely "curve fitting." Rather, it is the development of a set of numerical values for physical properties of this star system, such as the separation of the stars, their relative sizes, orbital eccentricity, and so on (see, for example, Wilson 1994b). Using these quantities, appropriate software generates the predicted light-curve for such a star system, in this case the program binarymaker3. The numerical "parameters" are manually adjusted to achieve a good fit of the predicted light curve to the observations.

Figure 3 shows the data used as the goal to be fit by the model, as well as the full cycle light curve predicted by the model. It consists of the data listed in Table 3, during primary and secondary eclipses, as well as some nights of out-of-eclipse observations that were used to compute the zeropoint shift between the observers, and establish flux = 1 for modeling.

Figure 3(A) shows all the data that were fit, and the final model.

Figure 3(B) is an enlargement around the primary eclipse. It shows that we have just one observing run through the primary eclipse (GB's data of RJD 7699). It is complemented by a pair of nights during ingress and egress by RF (7022 and 7114). A time of minimum was determined for the night of RJD 7699 (HJD 2457699.8739(2)) using the Kwee and van Woerden (1956) algorithm as implemented by the program AVE (Barberà 1996). A "synthetic" time of minimum was determined from RF's ingress and egress observations of RJDs 7022 and 7114 using "the digital tracing paper method." In this method, all the data are first "moved" to a single cycle by adding an integral number of periods to the time of each observation. The data are then plotted against the time away from a trial minimum, and then over-plotted with the same data time-reversed around the trial minimum. The trial time of minimum that gives the best visual match between the forward and reversed data is taken as the time of the eclipse—in this case, 2457059.269(2). The process is iterated if the resulting time of minimum implies a different period than was used to first move the observations to a single common cycle.

These two times of primary minima were used to determine the elements used to phase the data shown in Figure 3 (epoch 2457699.8739, period 18.3030 d). These elements perfectly "phase" the data in the primary eclipse (Figure 3(B)), but not the secondary eclipses (Figure 3(C)). The secondary eclipse data are grouped around two epochs. RF's data from RJDs 6958 and 6959 are just one night apart (epoch 2456958). GB's data from RJDs 7726–7782 span 56 days (epoch 2457763), centred 805 days after RF's data. These two epochs show a different phase for the secondary eclipse, and we fit them with two different models, differing only in the longitude of periastron.

From these data (plotted at larger scale) we observed that the primary eclipse is flat-bottomed, with duration 0.0022 of the cycle (0.0403 d), and flux 0.595 during the primary eclipse, and 0.655 during the secondary (the out-of-eclipse brightness defines flux of 1.0).

Times of minima were also estimated for two secondary eclipses corresponding to the aforementioned epochs RJD 6958 and 7763. Once again, "synthetic" minima were analysed using the digital tracing paper method, with the results listed in Table 4. The O–C (observed minus computed) values in Table 4 are the difference between observed eclipse times and the times predicted by a mean ephemeris with epoch 2457699.8739, period 18.3017, and a secondary eclipse phase of 0.5035. The deviations of the primary and eclipse times from a single linear ephemeris is shown in the O–C diagram of Figure 4.

To generate a modeled light curve, we must supply the modeling program with parameters that describe the two stars and their orbits. We start with initial estimates of these parameters and adjust them to improve the model fit.

Table 4. Times of minima for primary and secondary eclipses.

| Time of Minimum (RJD) | Cycle | Notes | O–C (days) |
|---|---|---|---|
| 1403.83(1) | –344 | (p), fit to NSVS and SuperWASP data by Furgoni (2014) | 0.26(1) |
| 7059.269(2) | –35 | (p), fit to RF observations, this paper | –0.045(2) |
| 7699.8739(2) | 0 | (p), GB observations, this paper | 0 |
| 1486.75(1) | –340 | (s), fit to NSVS and SuperWASP data by Furgoni (2014) | 0.24(1) |
| 6958.76(5) | –41 | (s), fit to RF observations, this paper | 0.041(5) |
| 7763.993(5) | 3 | (s), fit to GB observations, this paper | 0.001(5) |

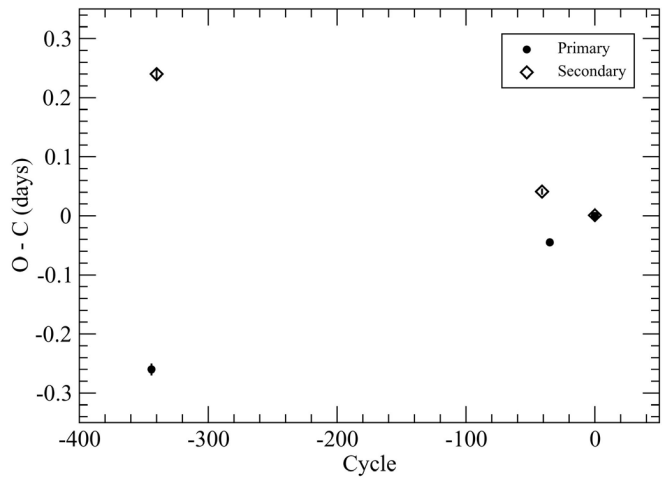

Figure 4. O–C ("observed minus calculated") diagram showing the observed eclipse times of Table 4 relative to time calculated using mean (of primary and secondary) elements (epoch 2457699.873, period 18.3017), and a secondary eclipse phase of 0.5035.

4.1. Initializing parameters: temperature

Effective temperature ($T_{eff}$) is a key parameter describing a star, but our modeling is based on only single color data, so we cannot tune stellar temperatures based on the model fit. In other words, the model is independent of stellar temperatures, and assumes both stars have the same $T_{eff}$. We use $T_{eff}$ = 22000, corresponding to the spectral type B2 and luminosity class V (Schmidt-Kaler 1982, p. 456). The spectral class used was taken from the *Catalogue of Stellar Spectral Classifications* (Skiff 2009–2016) compiled after a systematic review of the literature. In particular this spectral classification was originally provided by Voroshilov (Voroshilov *et al.* 1985). In any case, the spectral type considered will not affect our determination of the key orbital parameters of eccentricity and changing longitude of periastron.

4.2. Initializing parameters: geometric factors

Inspection of the light curve reveals some general properties of the system, as follows. During primary eclipse, a smaller star must be entirely behind, or transiting the face of a larger star (to give the flat-bottomed (total) eclipse). Inclination must be near 90° (to give the total primary eclipse), but different from 90° so as to produce the round-bottomed (partial) secondary eclipse. We must be observing an eccentric system with our line-of-sight nearly along its semi-major axis (longitude of periastron near 90°), so that the secondary eclipse occurs at near phase 0.50, and is of different duration than the primary. None of the data suggest brightness variations outside of the eclipses, so we expect a detached system with nearly spherical stars. With these qualitative guides as a starting point, the relative radii of the two stars, system eccentricity, and longitude of perihelion were determined by trial and error. The resulting parameters are listed in Table 5.

Table 5. Resulting model parameters for GSC 04052-01378.

| Parameter | Value |
|---|---|
| Mass ratio | Not determined[1] |
| Radius of star 1[2,3,7] | 0.0643(3)[7] |
| Radius of star 2[3,7] | 0.0788(4)[7] |
| Temperature of star 1 | 22000 K[4] |
| Temperature of star 2 | 22000 K[4] |
| Gravity brightening exponent of star 1 | 1.0[5] |
| Gravity brightening exponent of star 2 | 1.0[5] |
| Limb darkening coefficient for star 1 | 0.255[5] |
| Limb darkening coefficient for star 2 | 0.255[5] |
| Reflection coefficient for star 1 | 1.0[5] |
| Reflection coefficient for star 2 | 1.0[5] |
| Third light | 0.0[6] |
| Inclination | 88.77(5)°[7] |
| Longitude of periastron for epoch 245 7763 | 89.54(4)°[7] |
| Longitude of periastron for epoch 245 6958 | 88.82(4)°[7] |
| Eccentricity | 0.538(6)[7] |

1. Modeling results were insensitive to large variations in mass ratio. A value of 1.0 was used in modeling. Wilson (1994a, p. 930) states "for a detached binary…a light curve ordinarily carries insufficient information to fix the mass ratio reliably."
2. Star 1 is eclipsed during the primary eclipse.
3. The radii are $r_{back}$: "the radius of the star directed away from the other star, along the axis containing their mass centers," expressed as a fraction of the semimajor axis of the relative orbit of the two stars. See the BINARYMAKER3 documentation.
4. Temperatures are fixed, to correspond to the spectral type B2.
5. Values recommended by BINARYMAKER3 documentation, based on $T_{eff}$.
6. Assumed.
7. Adjusted to achieve model fit.

5. Results

Figure 3(A) shows the resulting model fit to the whole cycle, and 3(B) shows an enlargement at primary minimum. The modeled light curve is flat-bottomed in the primary eclipse, that is, star 1 passes entirely behind star 2 during this eclipse. Figure 3(C) shows an enlargement around the secondary eclipse (at the same horizontal scale as Figure 3(B)). The solid line is the fit to epoch 7763; the dashed line is a fit to epoch 6958. Only the longitude of periastron was changed to accomplish

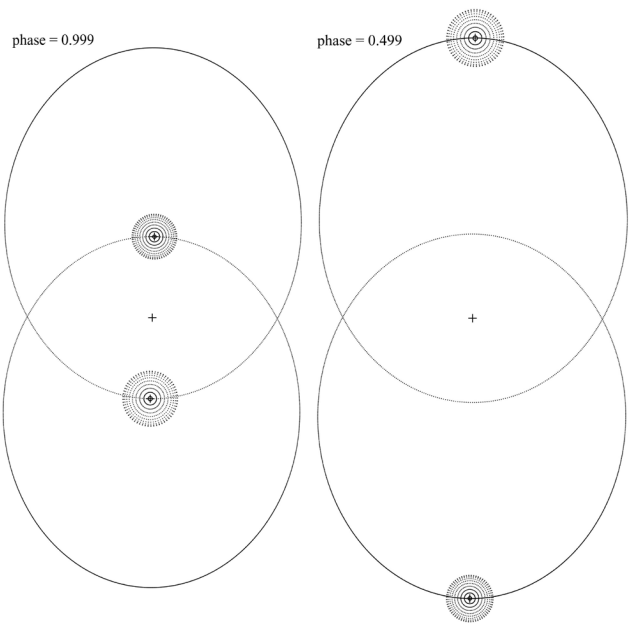

Figure 5. The binary system as viewed from above the plane of orbits, i.e. perpendicular to the line-of-sight from earth. Earth is towards the bottom of the page, in the plane of the page. The ellipses are the eccentric (non-circular) orbits of the stars. In the left image, the system is at primary eclipse, and the two stars are at their nearest approach to each other. In the right-hand image, the stars are at their maximum separation, and at secondary eclipse. When the stars are at maximum separation, i.e. maximum distance from the foci of their orbits, they move much more slowly—this is the cause of the greater duration of the secondary eclipse.

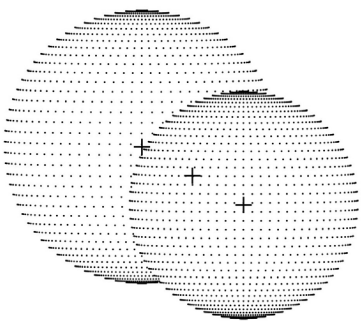

Figure 6. The binary system as viewed along the line-of-sight from earth, as the stars leave secondary eclipse. The smaller star is passing (from left to right) in front of the larger star, but the eclipse is partial (the smaller star does not cross directly in front of the larger star). In the primary eclipse (not shown), only the larger star is visible—the smaller star is hidden behind it. This is possible because the inclination is not exactly 90° and the orbit is eccentric (the stars are closer together at the time of primary eclipse).

this second fit from the first one, consistent with rotation of the line of apses.

The longitude of periastron changes 0.72(6)° between epochs 6958 and 7763 (Table 5); that is, in 44 cycles of the binary orbit. The rate of change is therefore $1.6(1) \times 10^{-2}$ °/orbit, $8.9(7) \times 10^{-4}$ °/day or $1.1(1) \times 10^{3}$ years/apsidal rotation. (See, for example, Martynov 1973.) This determination should be revisited once data over a longer time span are available—the observations used in this analysis span 17 years—just 1.5% of one rotation of the apse!

For planning future observations, the following elements are recommended. Note: the durations of the eclipses are approximately 10 hours and 31 hours, respectively.

Min I: HJDmin = 245 7699.8739 + 18.3030 × E

Min II: HJDmin = 245 7709.086 + 18.3009 × E

In Figures 5 and 6 we propose two different views of this high-eccentricity system: from above the plane of obits, perpendicular to the line-of-sight from Earth (Figure 5) and along the line-of-sight from Earth, as the stars leave secondary eclipse (Figure 6).

## 6. Conclusions

We only varied radii, inclination, eccentricity, and longitude of perihelion to make the model fit. The $T^{eff}$ used was adopted based on spectral type—it does not affect key results such as eccentricity, longitude of periastron, and apsidal motion. The gravity brightening exponent and limb-darkening coefficients were not varied from the values recommended for the spectral type of the star (see the BINARYMAKER3 documentation). Because we have no radial velocity data, and no out-of-eclipse brightness variation, we cannot derive stellar masses, nor the mass ratio, nor the absolute radii or separation of the two stars.

The resulting model parameters are listed in Table 5. BINARYMAKER3 is a forward-modeling program only: it does not adjust the parameters automatically, and provides no estimates of the uncertainty or standard error of the parameters. Nevertheless, in the process of manually performing trial-and-error fits, a one can observe how much of a change in a parameter makes a noticeable difference in the quality of fit, as assessed by visual inspection of residuals, and the computed sum of squared residuals between the observed light curve points and the fitted model. Once the parameters were determined, as reported in Table 5, perturbations were applied to one parameter at a time. A perturbation large enough to make a noticeable deterioration in the quality of fit was taken to be a 2-sigma error. The standard errors reported in Table 5 are half of those values. This process is, of course, subjective, and takes no account of the well-known correlation (non-independence) of the parameters of such models (Kallrath and Milone 2009, p. 174). It is likely that these standard errors significantly underestimate the difference between these parameters and those that might be derived from higher-quality multi-color photometry and radial velocity observations.

This star system's eccentricity of 0.538(6) identifies it as a high-eccentricity system. We have observed apsidal motion, and plan to observe more minima to permit a more confident analysis of the apsidal rotation rate. The system may also be showing a varying depth of secondary eclipse due to the apsidal rotation—this could be confirmed by monitoring with more precise (transformed) photometry. Finally, radial-velocity observations would permit determination of masses and absolute radii, thereby confirming and refining the spectral type.

**7. Acknowledgements**

This work has made use of the VizieR catalogue access tool, CDS, Strasbourg, France, and the International Variable Star Index (VSX) operated by the AAVSO, Cambridge, Massachusetts.

This work has made use of NSVS data obtained from the Sky Database for Objects in Time-Domain operated by the Los Alamos National Laboratory, and data from the DR1 of the WASP data (Butters *et al.* 2010) as provided by the WASP consortium, and the computing and storage facilities at the CERIT Scientific Cloud, reg. no. CZ.1.05/3.2.00/08.0144 which is operated by Masaryk University, Czech Republic.